\def\ptitle{Functional inversion for potentials in QM}
\input psfig.sty
\nopagenumbers
\magnification=\magstep1
\hsize 6.0 true in 
\hoffset 0.25 true in 
\emergencystretch=0.6 in                 
\vfuzz 0.4 in                            
\hfuzz  0.4 in                           
\vglue 0.1true in
\mathsurround=2pt                        
\topskip=24pt                            
\def\nl{\noindent}                       
\def\np{\hfil\vfil\break}                
\def\title#1{\bigskip\noindent\bf #1 ~ \tr\smallskip} 
\font\tr=cmr10                          
\font\bf=cmbx10                         
\font\sl=cmsl10                         
\font\it=cmti10                         
\font\trbig=cmbx10 scaled 1500          
\font\tiny=cmr8                         
\def\ma#1{\hbox{\vrule #1}}             
\def\ng{>\kern -9pt|\kern 9pt}          
\def\bra{{\rm <}}                       
\def\ket{{\rm >}}                       
\def\hi#1#2{$#1$\kern -2pt-#2}          
\def\hy#1#2{#1-\kern -2pt$#2$}          


\output={\shipout\vbox{\makeheadline
                                      \ifnum\the\pageno>1 {\hrule}  \fi 
                                      {\pagebody}   
                                      \makefootline}
                   \advancepageno}

\headline{\noindent {\ifnum\the\pageno>1 
                                   {\tiny \ptitle\hfil page~\the\pageno}\fi}}
\footline{}
\newcount\zz  \zz=0  
\newcount\q   
\newcount\qq    \qq=0  

\def\pref #1#2#3#4#5{\frenchspacing \global \advance \q by 1     
    \edef#1{\the\q}
       {\ifnum \zz=1 { %
         \item{[\the\q]} 
         {#2} {\bf #3},{ #4.}{~#5}\medskip} \fi}}

\def\bref #1#2#3#4#5{\frenchspacing \global \advance \q by 1     
    \edef#1{\the\q}
    {\ifnum \zz=1 { %
       \item{[\the\q]} 
       {#2}, {\it #3} {(#4).}{~#5}\medskip} \fi}}

\def\gref #1#2{\frenchspacing \global \advance \q by 1  
    \edef#1{\the\q}
    {\ifnum \zz=1 { %
       \item{[\the\q]} 
       {#2}\medskip} \fi}}

 \def\sref #1{~[#1]}

\def\references#1{\zz=#1
   \parskip=2pt plus 1pt   
   {\ifnum \zz=1 {\noindent \bf References \medskip} \fi} \q=\qq

\pref{\halld}{R. L. Hall, J. Phys. A:Math. Gen}{28}{1771 (1995)}{} 
\pref{\hallc}{R. L. Hall, Phys. Rev. A}{50}{2876 (1995)}{}         
\pref{\chada}{K. Chadan, C. R. Acad. Sci. Paris S\`er. II}{299}{271 (1984)}{}
\pref{\chadb}{K. Chadan and H. Grosse, C. R. Acad. Sci. Paris S\`er. II}{299}{1305 (1984)}{}
\pref{\chadc}{K. Chadan and R. Kobayashi, C. R. Acad. Sci. Paris S\`er. II}{303}{329 (1986)}{}
\pref{\chadd}{K. Chadan and M. Musette, C. R. Acad. Sci. Paris S\`er. II}{305}{1409 (1987)}{}
\bref{\chad}{K. Chadan and P. C. Sabatier}{Inverse Problems in Quantum Scattering Theory}{Springer, New York, 1989}{The `inverse problem in the coupling constant' is discussed on p406}
\bref{\zak}{B. N. Zakhariev and A.A.Suzko}{Direct and Inverse Problems: Potentials in Quantum Scattering Theory}{Springer, Berlin, 1990}{The `inverse problem in the coupling constant' is mentioned on p53}
\pref{\halla}{R. L. Hall, Phys. Rev A}{51}{1787 (1995)}{}          
\pref{\halle}{R. L. Hall, J. Math. Phys.}{40}{669 (1999)}{} 
\pref{\hallh}{R. L. Hall, J. Math. Phys.}{40}{2254 (1999)}{} 
\bref{\gel}{I. M. Gelfand and S. V. Fomin}{Calculus of Variations}{Prentice-Hall, Englewood Cliffs, 1963}{Legendre transformations are discussed on p 72.} 
\pref{\hallf}{R. L. Hall, J. Math. Phys.}{25}{2708 (1984)}{}       
\pref{\hallg}{R. L. Hall, J. Math. Phys.}{34}{2779 (1993)}{}       
 }

 \references{0}    

\vskip 1.0true in
\centerline{\trbig Functional Inversion for Potentials}
\vskip 0.1true in
\centerline{\trbig in Quantum Mechanics}
\vskip 0.5true in
\tr 
\baselineskip 12 true pt 
\centerline{\bf Richard L. Hall}\medskip
\centerline{\sl Department of Mathematics and Statistics,}
\centerline{\sl Concordia University,}
\centerline{\sl 1455 de Maisonneuve Boulevard West,}
\centerline{\sl Montr\'eal, Qu\'ebec, Canada H3G 1M8.}
\centerline{email:\sl~~rhall@cicma.concordia.ca}
\bigskip\bigskip

\parskip=5pt plus 1pt      
\baselineskip = 18true pt  
\centerline{\bf Abstract}\medskip
Let $E = F(v)$ be the ground-state eigenvalue of the Schr\"odinger Hamiltonian $H = -\Delta + vf(x),$ where the potential shape $f(x)$ is symmetric and monotone increasing for $x > 0,$ and the coupling parameter $v$ is positive.  If the {\it kinetic potential} $\bar{f}(s)$ associated with $f(x)$ is defined by the transformation $\bar{f}(s) = F'(v), \quad s = F(v)-vF'(v),$ then $f$ can be reconstructed from $F$ by the sequence $f^{[n+1]} = \bar{f}\circ\bar{f}^{[n]^{-1}}\circ f^{[n]}.$  Convergence is proved for special classes of potential shape; for other test cases it is demonstrated numerically.  The seed potential shape $f^{[0]}$ need not be `close' to the limit $f.$

\medskip\noindent PACS~~03 65 Ge

\np
  \title{1.~~Introduction}
We consider the Schr\"odinger operator
 
$$H = -\Delta + vf(x)\eqno{(1.1)}$$

\nl defined on some suitable domain in $L^{2}(\ma{R}).$  The potential has two aspects: an `attractive' potential shape $f(x),$ and a coupling parameter $v > 0.$  We assume that $f(x)$ is symmetric, non-constant, and monotone increasing for $x > 0.$ Elementary trial functions can be designed for such an operator to prove that for each $v > 0$ there is a discrete eigenvalue $E = F(v)$ at the bottom of the spectrum.  If, in addition to the the above properties, we also assume that $f$ is continuous at $x = 0$ and that it is piecewise analytic, then we are able to  prove\sref{\halld} that $f$ is uniquely determined by $F.$ The subject of the present paper is the reconstruction of the potential shape $f$ from knowledge of the `energy trajectory' $F.$   This is an example of what we  call `geometric spectral inversion'\sref{\halld, \hallc}.

Geometric spectral inversion should be distinguished from the `inverse problem in the coupling constant' which has been analysed in detail by Chadan {\it et al}\sref{\chada-\zak}.  In the latter problem the discrete part of the input data consists of the set $\{v_{i}\}$ of values of the coupling that yield a given fixed energy $E.$  Inversion from excited-state energy trajectories $F_{k}(v),\quad k > 0,$ has also been studied, by a complete inversion of the WKB approximation for bound states\sref{\halla}.   For the ground-state energy trajectory $F(v) = F_{0}(v)$ a constructive numerical inversion algorithm has been devised\sref{\halle}, and an inversion inequality has been established\sref{\hallh}. The work reported in the present paper also concerns inversion from the ground-state energy trajectory, but the approach uses functional methods which have a natural extension to the higher energy trajectories. 

Geometry is involved with this problem because we deal with a family of operators depending on a continuous parameter $v.$  This immediately leads to a family of spectral manifolds, and, more particularly, to the consideration of smooth transformations of potentials, and to the transformations which they in turn induce on the spectral manifolds.   This is the environment in which we are able to construct the following functonal inversion sequence that is the central theme of the present paper:

$$f^{[n+1]} = \bar{f}\circ\bar{f}^{[n]^{-1}}\circ f^{[n]} \equiv \bar{f}\circ K^{[n]}.\eqno{(1.2)}$$

\nl A {\it kinetic potential} is the constrained mean value of the potential shape $\bar{f}(s) = \bra f\ket,$ where the corresponding mean kinetic energy $s = \bra -\Delta\ket$ is held constant.  It turns out that kinetic potentials may be obtained from the corresponding energy trajectory $F$ by what is essentially a Legendre transformation\sref{\gel} $\bar{f} \leftrightarrow F$ given\sref{\hallf} by

$$\{\bar{f}(s) = F'(v),\quad s = F(v)-vF'(v)\}\quad\leftrightarrow\quad
\{F(v)/v = \bar{f}(s) - s\bar{f}'(s),\quad 1/v = - \bar{f}'(s)\}.\eqno{(1.3)}$$

\nl As we shall explain in more detail in Section 2, these transformations are well defined because of the definite convexities of $F$ and $\bar{f};$  they complete the definition of the inversion sequence (1.2), up to the choice of a starting seed potential $f^{[0]}(x).$  They differ from  Legendre transformations only because of our choice of signs.  The choice has been made so that the eigenvalue can be written (exactly) in the semi-classical forms

$$E = F(v) = \min_{s > 0}\left\{s + v\bar{f}(s)\right\} = \min_{x > 0}\left\{K^{[f]}(x) + vf(x)\right\}\eqno{(1.4)}$$

\nl where the kinetic- and potential-energy terms have the `usual' signs.  

After more than 70 years of QM (and even more of the Sturm-Liouville problem) it may appear to be an extravagance to seek to rewrite the min-max characterization of the spectrum in slightly different forms, with kinetic potentials and K functions.  The main reason for our doing this is that the new representations allows us to tackle the following problem: if $g$ is a smooth transformation, and we know the spectrum of $-\Delta + vf^{[0]}(x),$  what is the spectrum of $-\Delta + vg(f^{[0]}(x))$? In the forward direction (obtaining eigenvalues corresponding to a given potential), an approximation called the `envelope method' has been developed\sref{\hallf, \hallg}. The inversion sequence (1.2) was arrived at by an inversion of envelope theory, yielding a sequence of approximations for an initially unknown transformation $g$ satisfying $f(x) = g(f^{[0]}(x)).$

In order to make this paper essentially self contained, the representation apparatus is outlined in Section 2.  In Section 3 we use envelope theory to generate the inversion sequence.  In Section 4 it is proved that the energy trajectory for a pure power potential is inverted from an arbitrary pure-power seed in only two steps: thus $f^{[2]} = f$ in these cases. In Section 5 we consider the exactly soluble problem of the sech-squared potential $f(x) = -{\rm sech}^{2}(x).$  Starting from the seed $f^{[0]}(x) = -1 + x^{2}/20,$ we are able to construct the first iteration $f^{[1]}$ exactly; we then continue the sequence by using numerical methods. This illustration is interesting because the seed potential shape  $f^{[0]}$ is very different from that of the target $f(x)$ and has a completely different, entirely discrete, spectrum.   We consider also another sequence in which the seed is $f^{[0]}(x) = -1/(1+x/5).$  Convergence, which, of course, cannot be proved with the aid of a computer, is strongly indicated by both of these examples.
  \title{2.~~Kinetic potentials and K functions}
The term `kinetic potential' is short for `minimum mean iso-kinetic potential'.  If the Hamiltonian is $H = -\Delta + vf(x),$ where $f(x)$ is potential {\it shape,} and ${\cal D}(H) \subset L^{2}(\ma{R})$ is the domain of $H,$ then the ground-state kinetic potential $\bar{f}(s) = \bar{f}_{0}(s)$ is defined\sref{\hallf, \hallg} by the expression

$$\bar{f}(s) = \inf_{{{\scriptstyle \psi \in {\cal D}(H)} \atop {\scriptstyle (\psi,\psi) = 1}} \atop {\scriptstyle (\psi, -\Delta\psi) = s}} (\psi, f\psi).\eqno{(2.1)}$$

\nl The extension of this definition to the higher discrete eigenvalues (for $v$ sufficiently large) is straightforward\sref{\hallf} but not explicitely needed in the present paper. The idea is that the min-max computation of the discrete eigenvalues is carried out in two stages: in the first stage (2.1) the mean potential shape is found for each fixed value of the mean kinetic energy $s;$ in the second and final stage we minimize over $s.$  Thus we have arrive at the semi-classical expression which is the first equality of Eq.(1.4).  It is well known that $F(v)$ is concave ($F''(v) < 0$) and it follows immediately that $\bar{f}(s)$ is convex.  More particularly, we have\sref{\hallc} 

$$F''(v)\bar{f}''(s) = -{1 \over {v^{3}}}.\eqno{(2.2)}$$

\nl Thus, although kinetic potentials are {\it defined} by (2.1), the transformations (1.3) may be used in practice to go back and forth between $F$ and $\bar{f}.$

Kinetic potentials have been used to study smooth transformations of potentials and also linear combinations.  The present work is an application of the first kind. Our goal is to devise a method of searching for a transformation $g,$ which would convert the initial seed potential $f^{[0]}(x)$ into the (unknown) goal $f(x) = g(f^{[0]}).$  We shall summarize briefly how one proceeds in the forward direction, to approximate $F,$ if we know $f(x).$  The $K$ functions are then introduced, by a change of variable, so that the potential $f(x)$ is exposed and can be extracted in a sequential inversion process.

In the forward direction we assume that the lowest eigenvalue $F^{[0]}(v)$ of $H^{[0]} = -\Delta + vf^{[0]}(x)$ is known for all $v > 0$ and we assume that $f(x)$ is given; hence, since the potentials are symmetric and monotone for $x > 0,$ we have defined the transformation function $g.$  `Tangential potentials' to $g(f^{[0]})$ have the form $a + bf^{[0]}(x),$ where the coefficients $a(t)$ and $b(t)$ depend on the point of contact $x = t$ of the tangential potential to the graph of $f(x).$  Each one of these tangential potentials generates an energy trajectory of the form ${\cal F}(v) = av + F^{[0]}(bv),$ and the {\it envelope} of this family (with respect to $t$) forms an approximation $F^{A}(v)$ to $F(v).$  If the transformation $g$ has definite convexity, then $F^{A}(v)$ will be either an upper or lower bound to $F(v).$ It turns out\sref{\hallf} that all the calculations implied by this envelope approximation can be summarized nicely by kinetic potentials. Thus the whole procedure just described corresponds exactly to the expression:

$$\bar{f} \approx \bar{f}^{A} = g\circ\bar{f}^{[0]},\eqno{(2.3)}$$

\nl with $\approx$ being replaced by an inequality in case $g$ has definite convexity. Once we have an approximation $\bar{f}^{A},$ we immediately recover the corresponding energy trajectory $F^{A}$ from the general minimization formula (1.4).  

The formulation that reveals the potential shape is obtained when we use $x$ instead of $s$ as the minimization parameter.  We achieve this by the following general definition of $x$ and of the $K$ function associated with $f:$

$$f(x) = \bar{f}(s),\quad K^{[f]}(x) = \bar{f}^{-1}(f(x)).\eqno{(2.4)}$$

\nl The monotonicity of $f(x)$ and of $\bar{f}$ guarantee that $x$ and $K$ are well defined. Since $\bar{f}^{-1}(f)$ is a convex function of $f,$ the second equality in (1.4) immediately follows\sref{\hallg}.  In terms of $K$ the envelope approximation (2.3) becomes simply

$$K^{[f]} \approx K^{\left[f^{[0]}\right]}.\eqno{(2.5)}$$

\nl Thus the envelope approximation involves the use of an approximate $K$ function that no longer depends on $f,$ and there is now the possibility that we can invert (1.4) to extract an approximation for the potential shape.

We end this summary by listing some specific results that we shall need.  First of all, the kinetic potentials and $K$ functions obey\sref{\hallf, \hallg} the following elementary shift and scaling laws:

$$f(x) \rightarrow a + bf(x/t) \Rightarrow  \left\{\bar{f}(s) \rightarrow a + b\bar{f}(st^{2}),\quad K^{[f]}(x) \rightarrow {1 \over {t^2}}K^{[f]}\left({x \over t}\right)\right\}.\eqno{(2.6)}$$

\nl Pure power potentials are important examples which have the following formulas:

$$f(x) = |x|^{q} \Rightarrow \left\{\bar{f}(s) = \left ({P \over {s^{1 \over 2}}}\right )^{q}, \quad
K(x) = \left ({P \over x}\right )^{2}\right \},\eqno{(2.7)}$$

\nl where, if the bottom of the spectrum of $-\Delta + |x|^{q}$ is $E(q),$ then the $P$ numbers are given\sref{\hallg} by the folowing expressions with $n = 0:$

$$P_{n}(q) = \left\vert E_{n}(q)\right\vert^{{(2+q)} \over {2q}}\left[{2 \over {2+q}}\right]^{1 \over q}\left[{{|q|} \over {2+q}}\right]^{1 \over 2}, \quad q \neq 0.\eqno{(2.8)}$$

\nl We have allowed for $q < 0$ and for higher eigenvalues since the formulas are essentially the same.  The $P_{n}(q)$ as functions of q are interesting in themselves\sref{\hallg}: they have been proved to be monotone increasing, they are probably concave, and $P_{n}(0)$ corresponds exactly to the $\log$ potential.  By contrast the $E_{n}(q)$ are not so smooth: for example, they have infinite slopes at $q = 0.$  But this is another story.  An important observation is that the $K$ functions for the pure powers are {\it all} of the form $(P(q)/x)^{2}$ and they are invariant with respect to both potential shifts and multipliers: thus $a + b|x|^{q}$ has the same $K$ function as does $|x|^{q}.$  For the harmonic oscillator $P_n(2) = (n+{1 \over 2})^{2}, \quad n = 0,1,2,\dots.$ Other specific examples may be found in the references cited. 

The last formulas we shall need are those for the ground state of the sech-squared potential:

$$f(x) = -{\rm sech}^{2}(x) \Rightarrow \left\{\bar{f}(s) = -{{2s} \over {(s + s^2)^{1 \over 2} + s}}, \quad K(x) = {\rm sinh}^{-2}(2x)\right\}.\eqno{(2.9)}$$

  \title{3.~~The inversion sequence}
The inversion sequence (1.2) is based on the following idea. The goal is to find a transformation $g$ so that $f = g\circ f^{[0]}.$  We choose a seed $f^{[0]},$ but, of course, $f$ is unknown.  In so far as the envelope approximation with $f^{[0]}$ as a basis is `good', then an approximation $g^{[1]}$ for $g$ would be given by $\bar{f} = g^{[1]}\circ\bar{f}^{[0]}.$ Thus we have

 $$g \approx g^{[1]} = \bar{f}\circ\bar{f}^{[0]^{-1}}.\eqno{(3.1)}$$

\nl Applying this approximate transformation to the seed we find:

$$f \approx f^{[1]} = g^{[1]}\circ f^{[0]} = \bar{f}\circ\bar{f}^{[0]^{-1}}\circ f^{[0]} = \bar{f}\circ K^{[0]}.\eqno{(3.2)}$$

\nl We now use $f^{[1]}$ as the basis for another envelope approximation, and, by repetition, we have the ansatz (1.2), that is to say

$$f^{[n+1]} = \bar{f}\circ\bar{f}^{[n]^{-1}}\circ f^{[n]} = \bar{f}\circ K^{[n]}.\eqno{(3.3)}$$

A useful practical device is to invert the second expression for $F$ given in (1.4) to obtain

$$K^{[f]}(x) = \max_{v > 0}\left\{F(v) - vf(x)\right\}.\eqno{(3.4)}$$

\nl The concavity of $F(v)$ explains the $\max$ in this inversion, which, as it stands, is exact.  In a situation where $f$ is unknown, we have $f$ on both sides and nothing can be done with this formal result.  However, in the inversion sequence which we are considering, (3.4) is extremely useful.  If we re-write (3.4) for stage [n] of the inversion sequence it becomes:

$$K^{[n]}(x) = \max_{v > 0}\left\{F^{[n]}(v) - vf^{[n]}(x)\right\}.\eqno{(3.5)}$$

\nl In this application, the current potential shape $f^{[n]}$ and consequently $F^{[n]}(v)$ can be found (by shooting methods) for each value of $v.$  The minimization can then be performed even without differentiation (for example, by using a Fibonacci search) and this is a much more effective method for $K^{[n]} = \bar{f}^{[n]^{-1}}\circ f^{[n]}$ than finding $\bar{f}^{[n]}(s),$ finding the functional inverse, and applying the result to $f^{[n]}.$   
    
  \title{4.~~Inversion for pure powers}
We now treat the case of pure-power potentials given by

$$f(x) = A + B|x|^{q}, \quad q > 0,\eqno{(4.1)}$$

\nl where  $A$ and $B > 0$ are arbitrary and fixed.  We shall prove that, starting from another pure power as a seed, the inversion sequence converges in just two steps. The exact energy trajectory $F(v)$ for the potential (4.1) is assumed known.  Hence, so is the exact kinetic potential given by (2.7) and the general scaling rule (2.6), that is to say  

$$\bar{f}(s) = A + B\left ({P(q) \over {s^{1 \over 2}}}\right )^{q}.\eqno{(4.2)}$$

\nl We now suppose that a pure power is also used as a seed, thus we have

$$f^{[0]}(x) = a + b|x|^{p}\quad \Rightarrow \quad K^{[0]}(x) = \left({{P(p)} \over x}\right)^{2},\eqno{(4.3)}$$

\nl where the parameters $a,\quad b > 0, \quad p > 0$ are arbitrary and fixed.  The first step of the inversion (1.4) therefore yields

$$f^{[1]}(x) = \left(\bar{f}\circ K^{[0]}\right)(x) = A + B\left({{P(q)|x|} \over {P(p)}}\right)^{q}.\eqno{(4.4)}$$

\nl The approximate potential $f^{[1]}(x)$ now has the correct $x$ power dependence but has the wrong multiplying factor.  Because of the invariance of the $K$ functions to multipliers, this error is completely corrected at the next step, yielding:

$$K^{[1]}(x) =  \left({{P(q)} \over x}\right)^{2}\quad\Rightarrow\quad f^{[2]}(x) = \left(\bar{f}\circ K^{[1]}\right)(x) = A + B|x|^q.\eqno{[4.5]}$$

\nl This establishes our claim that power potentials are inverted without error in exactly two steps. 

The implications of this result are a little wider than one might first suspect.  If the potential that is being reconstructed has the asymptotic form of a pure power for small or large $x,$ say, then we know that the inversion sequence will very quickly produce an accurate approximation for that part of the potential shape.  More generally, since the first step of the inversion process involves the construction of $K^{[0]},$ the general invariance property $K^{[a+bf]} = K^{[f]}$ given in (2.6) means that the seed potential $f^{[0]}$ may be chosen without special consideration to gross features of $f$ already arrived at by other methods. For example, the area (if the potential has area), or the starting value $f(0)$ need not be incorporated in $f^{[0]},$ say, by adjusting $a$ and $b.$      
  \title{5.~~More general examples}
We consider the problem of reconstructing the sech-squared potential $f(x) = -{\rm sech}^{2}(x).$  We assume that the corresponding exact energy trajectory $F(v)$ and, consequently, the kinetic potential $\bar{f}(s)$ are known.  Thus\sref{\hallg}:
$$f(x) = -{\rm sech}^{2}(x) \Rightarrow \left\{F(v) = -\left((v+{1 \over 4})^{1 \over 2} - {1 \over 2}\right)^{2},\quad\bar{f}(s) = -{{2s} \over {(s + s^2)^{1 \over 2} + s}}\right\}.\eqno{(5.1)}$$

\nl We study two seeds.  The first seed is essentially $x^2,$ but we use a scaled version of this for the purpose of illustration in Fig.(1).  Thus we have 

$$f^{[0]} = -1 + {{x^2} \over 20}\quad\Rightarrow\quad K^{[0]}(x) = {1 \over {4x^{2}}}\eqno{(5.1)}$$

\nl This potential generates the exact eigenvalue

$$F^{[0]}(v) = -v + \left({v \over {20}}\right)^{1 \over 2}\eqno{(5.2)}$$

\nl which, like the potential itself, is very different from that of the goal.  After the first iteration we obtain

$$f^{[1]}(x) = \bar{f}\left(K^{[0]}(x)\right) = -{2 \over {1 + (1+4x^{2})^{1 \over 2}}}.\eqno{(5.3)}$$

\nl A graph of this potential is shown as $f1$ in Fig.(1). In order to continue analytically we would need to solve the problem with Hamiltonian $H^{[1]} = -\Delta + vf^{[1]}(x)$ exactly to find an expression for $F^{[1]}(v).$  We know no way of doing this. However, it can be done numerically, with the aid of the inversion formula (3.5) for $K.$  The first 5 iterations shown in Fig.(1) suggest convergence of the series. 

As a second example we consider the initial potential given by

$$f^{[0]}(x) = -{1 \over {1 + |x|/5}}.\eqno{(5.4)}$$

\nl In this case none of the steps can be carried out exactly.  In Fig.(2) we show the first five iterations.  Again, convergence is indicated by this sequence, with considerable progress being made in the first step. 

The numerical technique needed to solve these problems is not the main point of the present work.  However, a few remarks about this are perhaps appropriate.  As we showed in Ref.[\halle], by looking at three large values of $v$ we can determine the best model of the form $f(x) = A + B|x|^{q}$ that fits the given $F(v)$ for small $x < x_{a}.$  We have used this method here with $v = 10000\times\{1,{1 \over 2},{1 \over 4}\}$ and $x_{a} = 0.2$ in {\it all} cases. As indicated above, the inversion (3.5) was used to determine each $K^{[n]}$ function from the corresponding $F^{[n]}.$  For all the graphs shown in the two figures, the range of $v$ still to be considered for $x > x_{a}$  turned out to be $0.0008 < v < 175.$  With 40 points beyond $x_{a}$ on each polygonal approximation for $f^{[n]}(x),$  a complete set of graphs for one figure took about $4{1 \over 2}$ minutes to compute with a program written in C++ on a PentiumPro running at 200MHz. The exact expression for $f^{[1]}$ arising from the harmonic-oscillator starting point was very useful for verifying the behaviour of the program.
 
\hfil\vfil\break 
  \title{6.~~Conclusion}
Progress has certainly been made with geometric spectral inversion. The results reported in this paper suggest strongly that in some suitable topology, the inversion sequence (1.4) converges to $f.$ The `natural' extension of (1.4) to excited states leads to the conjecture that {\it each} of the following sequences converges to $f:$

$$f^{[n+1]}_{k} = \bar{f}_{k}\circ\bar{f}^{[n]^{-1}}_{k}\circ f^{[n]}_{k} \equiv \bar{f}\circ K^{[n]}_{k},\quad k = 0,1,2,\dots\eqno{(6.1)}$$

\nl For the examples studied by the inversion of the WKB approximation, inversion improved rapidly as $k$ increased and the view of the problem became more `classical'.

If an energy trajectory $F(v)$ is derived from a potential which vanishes at large distances, is bounded,  and has area, it is straightforward\sref{\halld} to `normalize' the function $F(v)$ by scaling so that it corresponds to a potential with area $2$ and lowest point $-1.$ The graphs of $F_{0}(v)$ for normalized potentials with square-well, exponential, and sech-squared shapes look very similar: for small $v$ they are asymptotically like $-v^{2},$ and for large $v$ they satisfy $\lim_{v\rightarrow\infty}\{F_{0}(v)/v\} = -1.$  These asymptotic features are the same for {\it all} such normalized potentials. We now know that encoded in the details of these $F_{0}(v)$ curves for intermediate values of $v$ are complete recipes for the corresponding potentials.  If, as the WKB studies strongly suggest, the code could be unravelled for the excited states too, the situation would become very interesting. What this would mean is that, given any one energy trajectory $F_{k}(v),$ we could reconstruct from it the underlying potential shape $f(x)$ and then, by solving the problem in the forward direction, go on to find {\it all} the other trajectories $\{F_{j}(v)\}_{j\neq k},$ and {\it all} the scattering data.  For large $k$ this would imply that a classical view alone could determine the potential and hence all the quantum phenomena which it might generate.
   \title{Acknowledgment}
Partial financial support of this work under Grant No. GP3438 from the Natural Sciences and Engineering Research Council of Canada is gratefully acknowledged. 
\np
  \references{1}
\np

\hbox{\vbox{\psfig{figure=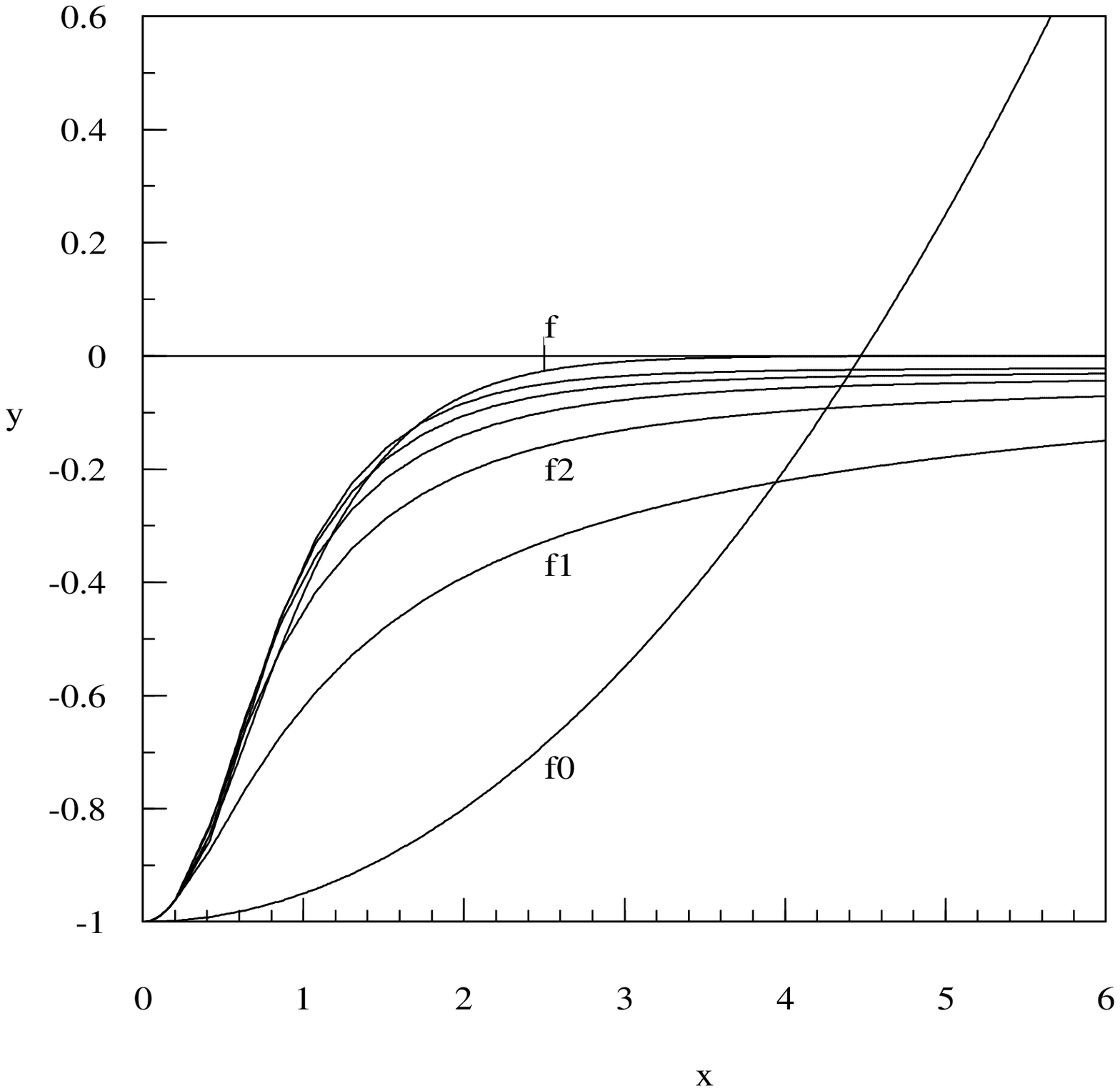,height=6in,width=6in,silent=}}}
\noindent {\bf Figure~(1)}~~The energy trajectory $F$ for the sech-squared potential $f(x) = -{\rm sech}^{2}(x)$ is approximately inverted starting from the seed $f^{[0]}(x) = -1 + x^{2}/20.$  The first step can be completed analytically yielding $f1 = f^{[1]}(x) = -2/\{1 + \sqrt{1 + 4x^{2}}\}.$ Four more steps $\{fk = f^{[k]}\}_{k=2}^{5}$ of the inversion sequence approaching $f$ are performed numerically.\medskip

\np
\hbox{\vbox{\psfig{figure=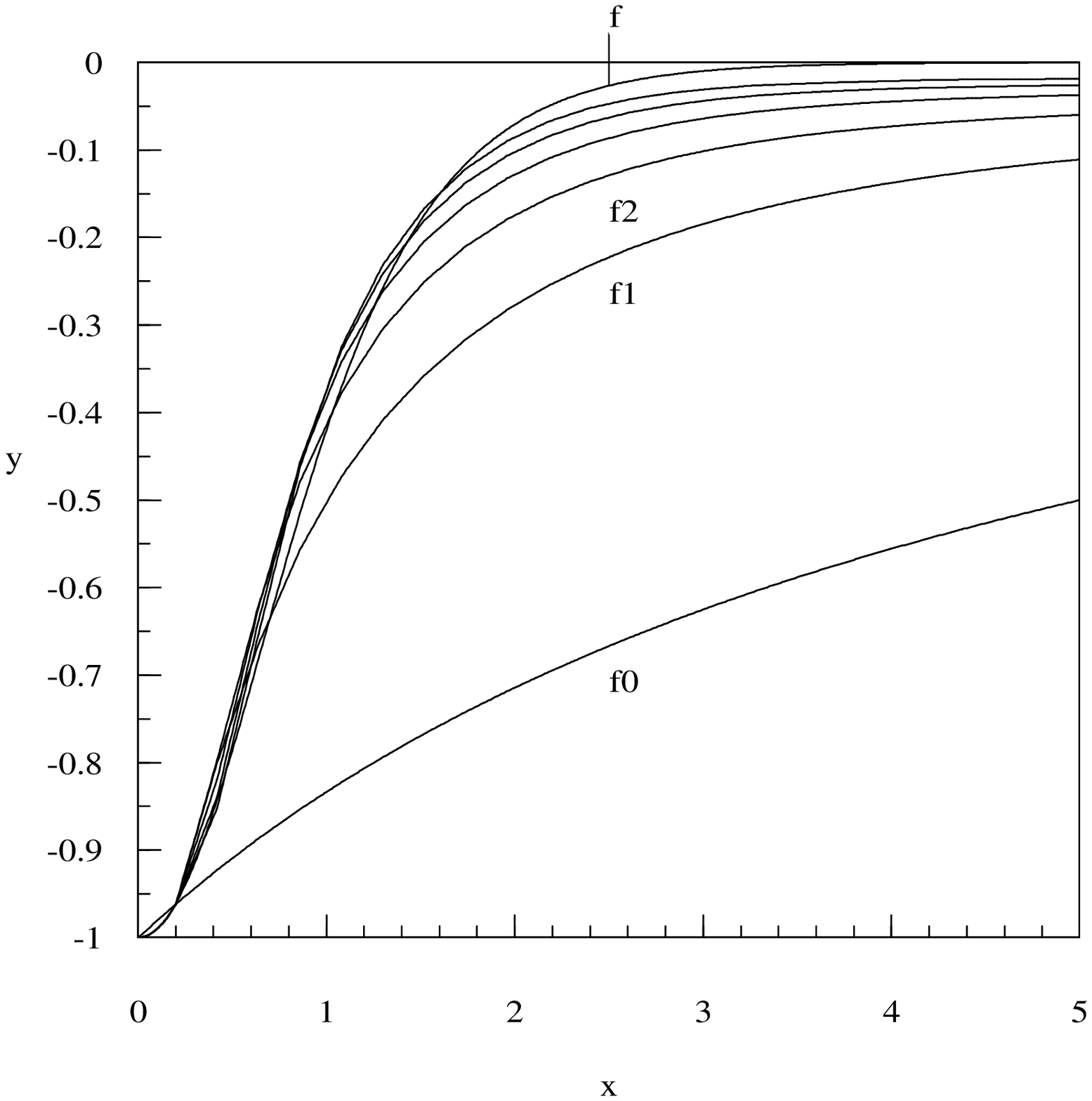,height=6in,width=6in,silent=}}}
\noindent{\bf Figure~(2)}~~The energy trajectory $F$ for the sech-squared potential $f(x) = -{\rm sech}^{2}(x)$ is approximately inverted starting from the seed $f0 = f^{[0]}(x) = -1/(1+x/5).$  The first 5 steps $\{fk = f^{[k]}\}_{k=1}^{5}$  of the inversion sequence approaching $f$ are performed numerically. \medskip

\hfil\vfil
\end